\documentclass[namedreferences]{SolarPhysics}
\usepackage[optionalrh,natbib]{spr-sola-addons} 
\usepackage{graphicx}                    
\usepackage{amssymb}                     
\usepackage{color}                       
\usepackage{url}                         

\newcommand{\degree}{^{\circ}}

\newcommand{\aap}{    {\it Astron. Astrophys.}}

\newcommand{\apj}{    {\it Astrophys. J.}}
\newcommand{\apjl}{    {\it Astrophys. J.}}

\newcommand{\grl}{    {\it Geophys. Res. Lett.}}

\newcommand{\solphys}{{\it Solar Phys.}}

\newcommand{\ssr}{    {\it Space Sci. Rev.}}

\begin{document}

\begin{article}

\begin{opening}

\title{Relation between the 3D-geometry of the coronal wave and associated CME during the 26 April 2008 event}

%
\author{M.~\surname{Temmer}$^{1}$\sep
       A.M.~\surname{Veronig}$^{1}$\sep
        N.~\surname{Gopalswamy}$^{2}$\sep
       S.~\surname{Yashiro}$^{2}$}

%

%
  \institute{$^{1}$ Kanzelh\"ohe Observatory-IGAM, Institute of Physics, University of Graz,
Universit\"atsplatz 5, A-8010 Graz, Austria \\  email: \url{manuela.temmer@uni-graz.at} \\
                     email: \url{asv@igam.uni-graz.at} \\
             $^{3}$ NASA Goddard Space Flight Center, Code 695, Greenbelt, MD 20771, USA \\
         email: \url{nat.gopalswamy@nasa.gov} \\
             }

\begin{abstract}
We study the kinematical characteristics and 3D geometry of a large-scale coronal wave that occurred in association with the 26 April 2008 flare-CME event. The wave was observed with the EUVI instruments aboard both STEREO spacecraft (STEREO-A and STEREO-B) with a mean speed of $\sim$240~km~s$^{-1}$. The wave is more pronounced in the eastern propagation direction, and is thus, better observable in STEREO-B images. From STEREO-B observations we derive two separate initiation centers for the wave, and their locations fit with the coronal dimming regions. Assuming a simple geometry of the wave we reconstruct its 3D nature from combined STEREO-A and STEREO-B observations. We find that the wave structure is asymmetric with an inclination towards East. The associated CME has a deprojected speed of $\sim$750$\pm$50~km~s$^{-1}$, and shows a non-radial outward motion towards the East with respect to the underlying source region location. Applying the forward fitting model developed by \cite{thernisien06}, we derive the CME flux rope position on the solar surface to be close to the dimming regions. We conclude that the expanding flanks of the CME most likely drive and shape the coronal wave.
\end{abstract}
%
\keywords{Shock waves, Coronal Mass Ejections}
\end{opening}

%
\section{Introduction}

Wave-like disturbances in the solar corona were for the first time imaged by the Extreme-ultraviolet Imaging Telescope \citep[EIT;][]{delaboudiniere95} instrument onboard the {\it Solar and Heliospheric Observatory} (SOHO), thereafter called EIT waves \citep{moses97,thompson98} or, more generally, EUV waves. Their generation mechanism and nature is still an issue of debate. In possible scenarios it is assumed that these disturbances are fast-mode MHD waves which are flare-initiated and/or CME driven \citep[e.g.][]{wills99,wang00,warmuth01,long08,veronig08, gopalswamy09}, solitons or slow-mode waves \citep[e.g.][]{wills-davey07}. Other models suppose that these disturbances are non-wave features that occur due to magnetic reconfiguration associated with the erupting CME \citep[e.g.][]{delanee00,chen02,attrill09,dai10}. Recently, theoretical models were developed that combine wave and non-wave scenarios within hybrid models \citep[see][]{zhukov04,cohen09,liu10}.

The original model developed by \cite{uchida68} assumes that the 3D structure of a shock wave in the corona sweeps over the chromosphere which would account for the observations of coronal as well as chromospheric wave signatures (Moreton waves imaged in H$\alpha$). Based on combined observations in different wavelengths of a coronal wave, \cite{narukage02} found from the resulting differences in propagation a 3D structure of the wave. A recent review by \cite{vrsnak08} describes the formation of MHD waves with a 3D piston mechanism driven either by the CME expansion or by a flare-associated pressure pulse. The {\it Solar Terrestrial Relations Observatory} (STEREO) provides new opportunities to observe coronal waves under different vantage points. Intriguing results giving evidence for the 3D geometry of coronal waves are reported for the STEREO quadrature event from 13 February 2009. For this event the EUV wave could be observed from the lateral direction as well as simultaneously on-disk. From these unique observations \cite{kienreich09} and \cite{patsourakos09} conclude that the disturbance is a 3D fast-mode MHD wave which is partly driven by the flanks of the associated CME. They further show that the EUV wave is propagating at heights of about 100~Mm above the solar surface which is comparable to the coronal scale height for quiet Sun conditions \citep{patsourakos09b}. In a recent study by \cite{veronig10} the full 3D dome of an EUV wave was identified, and interpreted to be freely propagating in the lateral direction after the lateral expansion of the CME stopped, whereas the upper part of the wave dome was permanently driven during the eruption. For recent reviews on the issue of EUV wave initiation and their nature we refer to \cite{wills10}, \cite{gallagher10}, and \cite{warmuth10}.

During the 26 April 2008 CME-flare event a coronal shock wave was imaged with the Extreme Ultraviolet Imager (EUVI) aboard both spacecraft of the STEREO mission, which were separated from each other by 49.5$\degree$. In this paper we present a study on the kinematics and geometry of the wave as seen simultaneously from two different vantage points. These observations are combined with a forward fitting model of the erupting flux rope, which enables us to derive geometrical information about the 3D structure of the disturbance.

\begin{figure}
 \centerline{\includegraphics[width=0.98\textwidth,clip=]{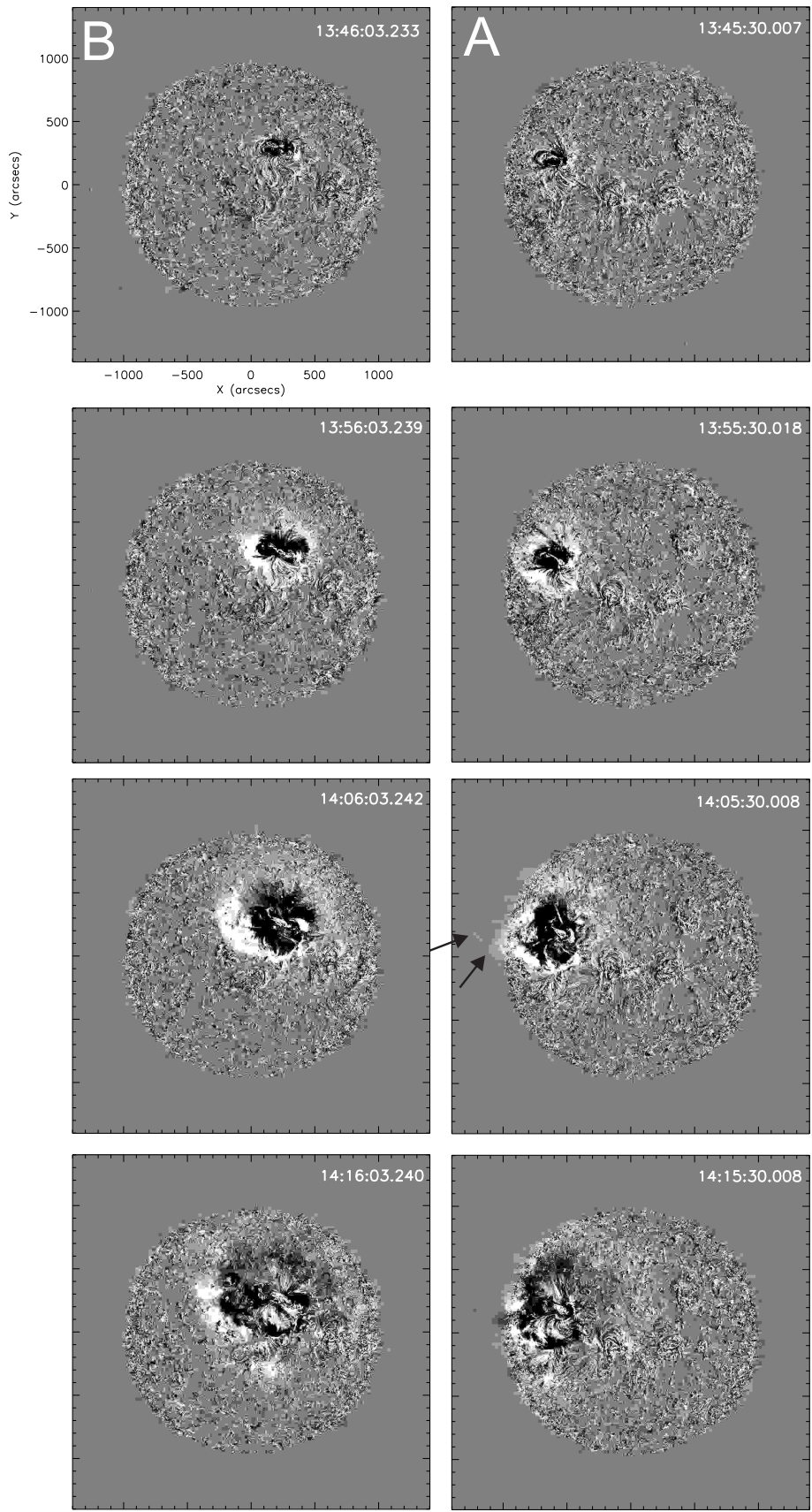}}
 \caption{Sequence of running difference images from STEREO-A and -B observations in the EUVI 195~{\AA} passband. A dome shaped structure observed from STEREO-A is indicated with arrows. See also the online animation.}
    \label{seq}
\end{figure}

\section{Data and Methods}

\begin{figure}
 \centerline{\includegraphics[width=1\textwidth,clip=]{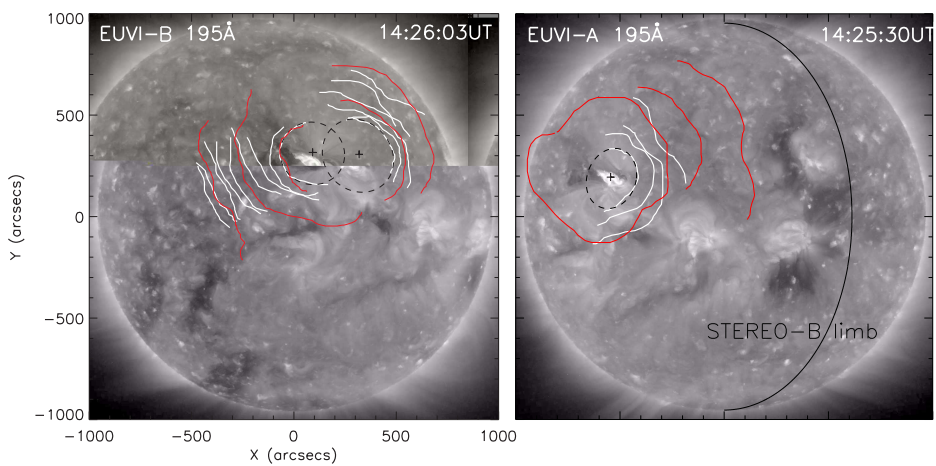}}
 \caption{Traced wave fronts observed in STEREO-B (left) and STEREO-A (right) EUVI 195~\AA~(red lines) and 171~\AA~(white lines) images. Black dashed circles indicate the fit to the first wave front as seen in 171~{\AA}. Black crosses give the estimated initiation center of the wave. The limb of STEREO-B is indicated in the image of STEREO-A.}
    \label{waveAB}
\end{figure}

The EUV wave under study occurred on 26 April 2008 and was associated with a CME and a flare of GOES class B3.8/H$\alpha$ importance SF from a source region located at N08E08 (Earth view). GOES soft X-ray (SXR) observations reveal that the thermal emission of the flare starts at 13:54~UT with a maximum at 14:08~UT. The event could be observed from both STEREO satellites (STEREO-A and STEREO-B) which were separated by 49.5$\degree$. Complementary observations from SOHO/LASCO and EIT provide information from Earth view. We stress that the EUV wave under study was associated with a coronal\footnote{\url{ftp://ftp.ngdc.noaa.gov/STP/SOLAR_DATA/SOLAR_RADIO/BURSTS/}} and interplanetary type II radio burst observed by Wind/WAVES\footnote{\url{http://cdaw.gsfc.nasa.gov/CME_list/radio/waves_type2.html}}. The type II burst in the metric and interplanetary space were studied in \cite{gopalswamy09b}.

The EUVI instrument and the coronagraphs COR1 and COR2 are part of the Sun Earth Connection Coronal and Heliospheric Investigation
\citep[SECCHI;][]{howard-stereo08} instrument suite onboard the STEREO mission \citep{kaiser08}, launched in October 2006. The STEREO mission consists of two identical spacecraft, which orbit the Sun ahead \mbox{(STEREO-A)} and behind \mbox{(STEREO-B)} the Earth near the ecliptic plane, and slowly separate from each other by $\sim$44$^\circ$ per year. EUVI observes the chromosphere and low corona in four different EUV emission lines out to 1.7\,$R_s$ (with $R_s$ the solar radius) \citep{wuelser04,howard-stereo08}. During the coronal wave event under study, the EUVI imaging cadence was 2.5 min in the 171~{\AA} and 10 min in the 195~{\AA} filter. For the comparative analysis of the early evolution of the associated CME in the low corona we use data from the STEREO/SECCHI inner coronagraph COR1, a classic Lyot internally occulting refractive coronagraph with a field-of-view (FOV) from 1.4 to 4\,$R_s$, and COR2 the externally occulted coronagraph with a FOV of 2 to 15\,$R_s$ \citep{howard-stereo08}. For our study we used polarized brightness COR2 images. In addition, SOHO/LASCO C2 (FOV: 1.5\,$R_s$--6\,$R_s$) and C3 (FOV: 3.5\,$R_s$--30\,$R_s$) data are studied \citep{brueckner95}.

The coronal wave is tracked by manually detecting the wave fronts, separately for SECCHI/EUVI-A and -B observations in both the 195~{\AA} and 171~{\AA} passband, respectively. From a circular fit to the first wave front, the initiation center of the wave is derived \citep[for more details see][]{veronig06mor}. Figure~\ref{waveAB} shows direct images of EUVI-A and EUVI-B in the 195~{\AA} passband together with the wave fronts determined from EUVI 171~{\AA} and 195~{\AA} running difference images. The distance of the wave to the initiation center is calculated by averaging the measured distance of each point of the tracked wave front from the derived initiation center along the spherical solar surface.

In order to infer the direction of motion of the CME we apply the triangulation method developed by \cite{temmer09b} on the distance-time measurements of the leading edge of the CME observed from LASCO, STEREO-A, and -B. This method uses the information of the spacecraft separation angles and takes the propagation direction of the CME as a free parameter, in order to transform SOHO/LASCO distance-time measurements to STEREO-A and STEREO-B view. By comparing the transformed distances with the distances actually observed from STEREO, the direction of motion, which gives the minimum deviation, results in the best estimation of the "true" CME propagation direction. To obtain an independent result for the propagation direction of the CME as well as to derive the surface location of the CME flux rope we also use the flux rope forward fitting model developed by \cite{thernisien06} and \cite{thernisien09}. This model is a raytrace simulation method which enables us to compute synthetic total and polarized brightness images using the Thomson scattering formulae from an assumed electron density model. Since the graduated cylindrical shell (GCS) model is a reasonable simulation of a flux-rope CME it can be applied to investigate the appearance of a CME. Characteristic CME parameters (width, cone angle, latitude, longitude) are derived by fitting the density model until a best match is found for contemporaneous image pairs from STEREO-A and STEREO-B, which observe the CME from two different vantage points.

\section{Results}

\begin{figure}
 \centerline{\includegraphics[width=0.8\textwidth,clip=]{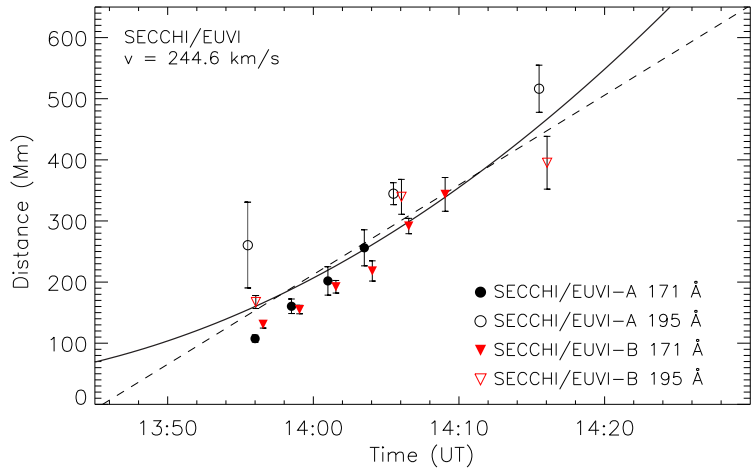}}
 \caption{Wave kinematics for the western direction as derived from EUVI 171~\AA~and 195~\AA~images of STEREO-A and STEREO-B (cf.~Figure~\ref{waveAB}). Solid and dashed lines show the quadratic and linear fit, respectively, to the data points. The mean velocity, $v$, resulting from the linear fit is given in the legend.}
    \label{kin1}
\end{figure}

Figure~\ref{seq} shows a sequence of EUVI 195~\AA~running difference images from STEREO-A and -B (see also the online movie). The location of the AR from which the wave is launched is E08 (if not stated otherwise, heliographic coordinates refer to Earth view). Hence, from the vantage point of STEREO-B the region is located less than 15$\degree$ west from Sun center, whereas from the vantage point of STEREO-A it is more than 35$\degree$ East. Therefore, measuring the wave fronts from STEREO-B is less affected by projection effects originating from the three-dimensional nature of the wave \citep[see][]{kienreich09,patsourakos09b}. Both STEREO-A and -B observations reveal that the wave intensity is higher for the eastern propagation direction. The wave appears to be brighter as well as more diffuse in 195~{\AA} compared to the 171~{\AA} passband. In general, coronal disturbances are found to be better observed in 195~{\AA} than in other wavelengths \citep{wills99}. We also note that a dome shaped structure is identified in STEREO-A observations which images the disturbance from an almost lateral direction.

Figure~\ref{waveAB} shows all tracked wave fronts for each spacecraft extracted from images in the EUVI 195~{\AA} as well as 171~{\AA} filter. From the vantage point of STEREO-A the AR is close to the eastern limb, therefore, the wave is best observed for the western direction (although less intense) and we focused in the construction of the wave kinematics to this direction. STEREO-B imagery enable us to track the wave in both the eastern and western propagation direction. By fitting circles to the first wave fronts separately for the eastern and western propagation direction observed in STEREO-B, two initiation centers on opposite sides of the AR are derived. This is obtained from EUVI-B 195~{\AA} as well as 171~{\AA} observations.

Figure~\ref{kin1} shows the derived wave kinematics for the western propagation direction obtained from STEREO-A and -B. In the 171~{\AA} passband the wave can be tracked over a shorter distance than in 195~{\AA} due to the weaker signal. For this propagation direction distances as derived from the 195~{\AA} filter deviate from those measured in 171~{\AA}. This is found from STEREO-A as well as STEREO-B measurements. Especially the first front observed in STEREO-A 195~{\AA} differs clearly from a circular shape since part of the wave is detected off limb. Performing a linear fit over all measurements we obtain a mean velocity of $\sim$245$\pm$20~km~s$^{-1}$. The (point-like) center of initiation from STEREO-A 171~{\AA} observations is derived at [$-$544$''$,193$''$], from STEREO-B 171~{\AA} at [318$''$,306$''$] from Sun center. A transformation of the coordinates from A to B view gives [253$''$,258$''$], i.e. there is a mismatch of $\approx$50$''$ when deriving the wave initiation center from the different vantage points. We assume that this difference is due to projection effects, uncertainties in the manual tracking of the wave as well as due to the simplified assumption of a point-like center of initiation \citep[see e.g.][]{muhr10}.

\begin{figure}
 \centerline{\includegraphics[width=0.8\textwidth,clip=]{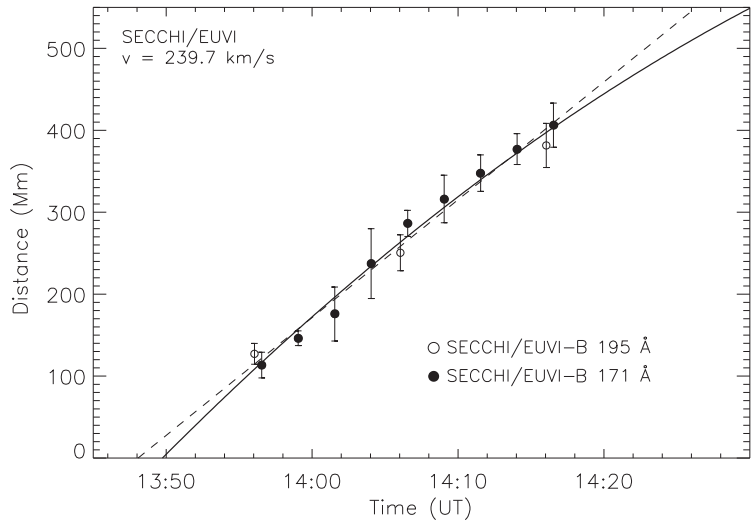}}
 \caption{Wave kinematics for the eastern direction as derived from EUVI 171~\AA~and 195~\AA~images observed with STEREO-B. Solid and dashed lines show the quadratic and linear fit, respectively, to the data points. The mean velocity, $v$, resulting from the linear fit to the distance-time measurements is given in the legend.}
    \label{kin0}
\end{figure}

Figure~\ref{kin0} shows the derived wave kinematics from STEREO-B observations for the eastern propagation direction. The mean velocity of the wave is $\sim$240$\pm$12~km~s$^{-1}$ which is comparable to the western propagation direction. The wave extracted from 171~{\AA} images can be tracked up to the same distance as in the 195~{\AA} passband.  Since for the western propagation direction we observe the wave well in STEREO-A, we can exclude that the wave ``disappears'' behind the STEREO-B limb. Therefore we infer an asymmetry in the intensity of the wave, i.e.\ the eastern propagation direction is more intense. In the following we will show that the main direction of the CME propagation as well as the 3D structure of the wave are in accordance with the more intense eastern wave propagation direction.

\subsection{Propagation direction of the associated CME}

\begin{figure}
 \centerline{\includegraphics[width=0.9\textwidth,clip=]{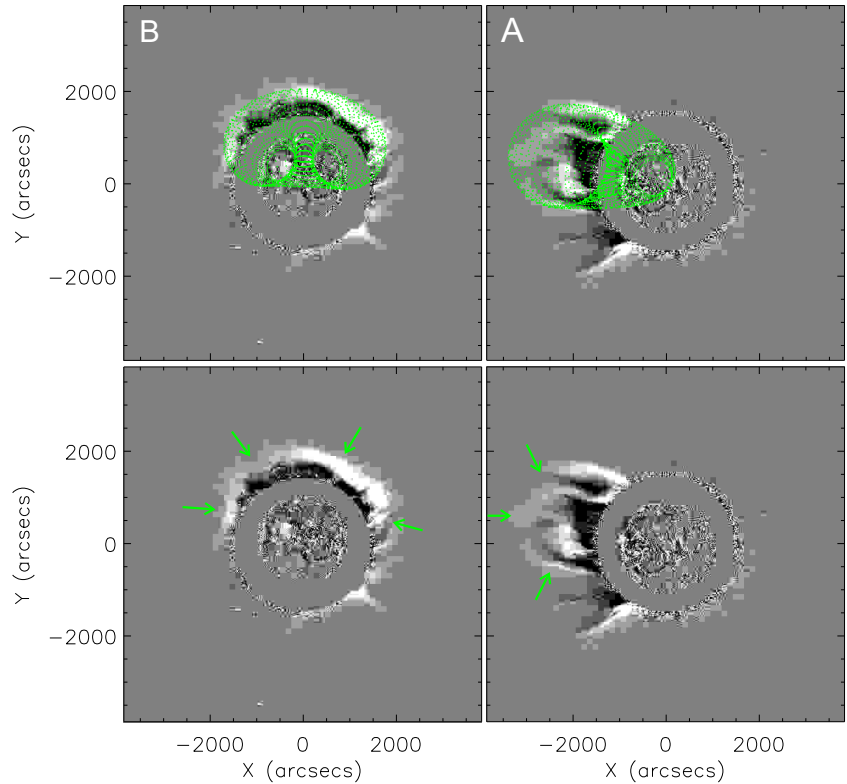}}
 \caption{Flux rope (green mesh) resulting from the raytracing-fitting method applied to COR1 observations from STEREO-A (right) and STEREO-B (left) overplotted on composite EUVI~195~\AA~-COR1 difference images from $\sim$14:45~UT. Arrows outline the shape of the CME to which the flux rope was fitted. See also the online animations of STEREO-A and -B composites from EUVI and COR1 running difference images.}
    \label{raytrace}
\end{figure}

\begin{figure}
 \centerline{\includegraphics[width=0.9\textwidth,clip=]{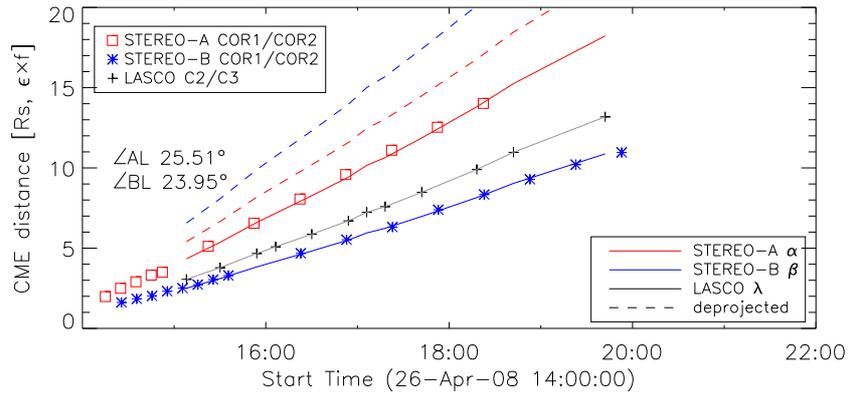}}
 \caption{Triangulation method from \cite{temmer09b} applied to the derived CME height-time data from observations of different
spacecraft. Solid lines are the modeled distances for STEREO-A view (red), {STEREO-B} view (blue), and LASCO (gray). The red and blue dashed lines show the resulting de-projected distances separately derived from LASCO/STEREO-A and LASCO/STEREO-B comparison. The separation angle between the
STEREO spacecraft with respect to LASCO is labeled ($\angle$AL, $\angle$BL).}
    \label{triang}
\end{figure}

\begin{figure}
\centerline{\includegraphics[width=0.8\textwidth,clip=]{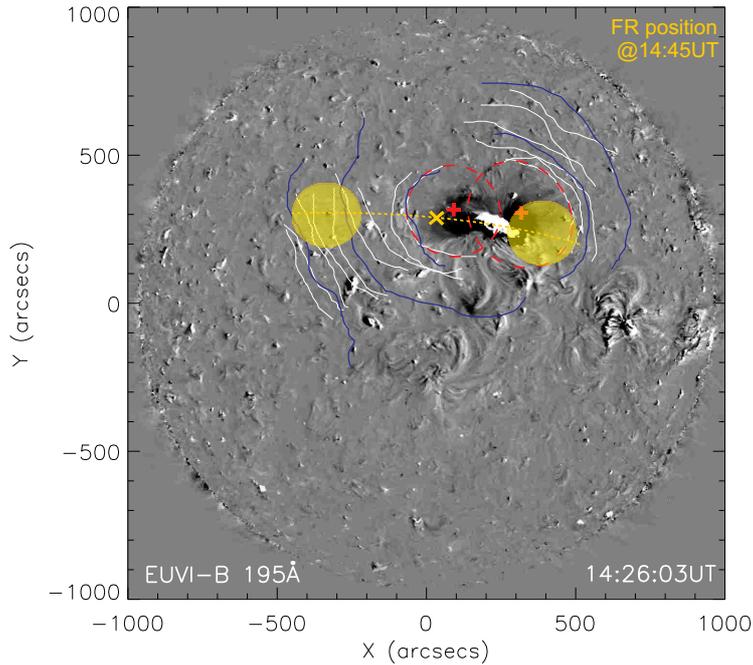}}
 \caption{Traced wave fronts (13:56UT--14:16UT) from STEREO-B EUVI 195\AA~in blue and 171\AA~in white plotted on a base difference image. Red dashed circles give the fit to the first wave front observed in EUVI 171\AA~from which the initiation centers (red crosses) of the wave are obtained within the bipolar dimming regions. Results from the raytracing tool derived from COR1 images at 14:45~UT (cf.~Figure~\ref{raytrace}) are indicated in yellow: shaded circles mark the position and extension of the footpoints of the flux rope, cross and dashed line give the apex and inclination of the flux rope projected on the solar surface.}
    \label{dimming}
\end{figure}

In Figure~\ref{raytrace} the result of the forward fit to the STEREO-A and -B COR1 image pair recorded at 14:45~UT is shown using the model by \cite{thernisien06} and \cite{thernisien09}. The CME observations from 14:45~UT do not overlap in time with the EUV wave, but are only $\sim$20~min after the last detected wave front. COR1 observations from STEREO-A and -B would show the CME in earlier images. However, the more of the CME structure is visible the more reliable is the forward fit. As can be seen in the lower right panel of Figure~\ref{raytrace}, the structure to the right of the top arrow possibly corresponds to a deflected streamer and not to the CME eruption itself (see also the accompanying movie). Therefore, it is possible that the CME fit, which includes this structure, overestimates the true width of the CME. From the fit we derive a CME propagation direction of E20--E25 and a width of $\sim$60$\degree$ as well as the position of the flux rope on the solar surface (cf.\ Figure~\ref{dimming}). Figure~\ref{triang} shows a second method from which the CME propagation direction is derived using the triangulation method developed by \cite{temmer09b}. From the triangulation of the STEREO-A and -B CME kinematics we obtain a propagation direction of E28--E35 (see Figure~\ref{triang}) which is consistent with the result derived from the flux rope forward fit. For comparison, the AR from which the CME is launched has a longitudinal position of E08. Using the derived information of the CME propagation direction, we obtain a deprojected CME speed of $\sim$750$\pm$50~km~s$^{-1}$.

Figure~\ref{dimming} shows a base difference image from STEREO-B EUVI 195~{\AA} and the derived wave fronts over the time range 13:56~UT--14:16~UT together with the flux rope position on the solar surface obtained with forward fitting applied to the COR1 STEREO-A and -B image pairs at 14:45~UT. The EUVI base difference image reveals coronal dimming regions on opposite sides of the AR. Their location is consistent with the two initiation centers obtained for the wave. The position of the footpoints of the flux rope obtained from the forward fit is also closely located to the dimming regions. The eastern footpoint of the flux rope is positioned to the east from the AR and outside the dimming region. The western footpoint is close to the AR and within the dimming area.

\subsection{3D geometry of the wave}

\begin{figure}
\centerline{\includegraphics[width=1\textwidth,clip=]{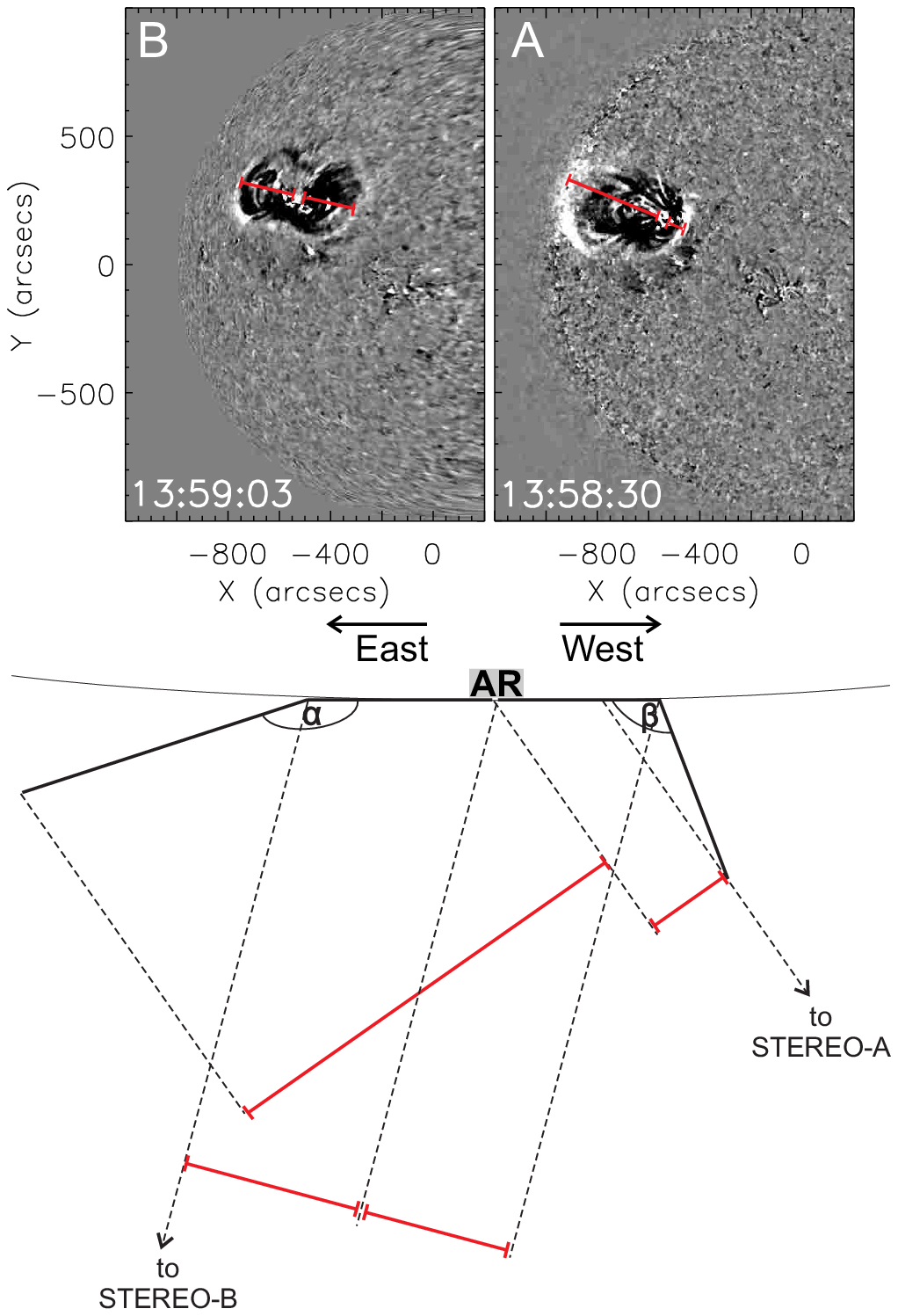}}
 \caption{Top: EUVI-A and -B 171\AA~difference images of the wave at 13:59~UT. Note that the STEREO-B image (left) is aligned to the viewing direction of STEREO-A (right) in order to make the differences in the morphology of the wave as viewed from two different vantage points clearly visible. The lateral extension of the wave is schematically indicated with red bars. Bottom: Reconstructed 3D geometry (view from solar north pole) with $\alpha\sim162\degree$ and $\beta\sim111\degree$. AR indicates the location of the active region. For details of the reconstruction see text.}
    \label{3d}
\end{figure}

Figure~\ref{3d} (top panels) shows difference images of the wave in EUVI 171~{\AA} as observed from STEREO-A and STEREO-B at 13:59~UT, respectively. For a better comparison the STEREO-B image is rotated to the viewing angle of STEREO-A. Imaging the wave from two different vantage points reveals differences in the morphology of the wave as well as ``foreshortening effects''. This hints at the 3D nature of the wave, and can be used to infer information of its 3D geometry. For the sake of simplicity we assume that the wave consists of a base and two lateral borders which are inclined with respect to the vertical. As can be seen from the red bars in Figure~\ref{3d} the distances between the active region and the eastern and western part of the wave are similar observed from STEREO-B, but different observed from STEREO-A. We therefore may assume that STEREO-B observes the base of the wave nearly free from projection effects. Using the observed extension of the eastern and western part of the wave as seen from STEREO-A, and knowing the location of the active region as well as the spacecraft positions, we can simply use trigonometric functions to derive the inclination of the borders of the wave. The results are schematically drawn in the bottom panel of Figure~\ref{3d}, revealing that the wave structure is asymmetric, i.e.\ more inclined towards East.

\section{Discussion and Conclusion}

For the EUV wave that occurred on 26 April 2008 we obtained two initiation centers located on either side of the AR from which the wave was launched. The two initiation centers are located within the coronal dimming regions. A similar result was found by \cite{muhr10} for the H$\alpha$ Moreton wave of 28 October 2003. For this event the initial location of the wave center was derived to be located close to the dimming regions well outside from the AR core, which are indicative of the footpoints of the flux rope of a CME \citep[e.g.][]{sterling97}. Model simulations of fast-mode MHD waves by \cite{wang00} showed that the wave disturbances needed to be launched in the periphery of the AR to expand horizontally over the surface, whereas the initiation in the core of the AR with strong magnetic fields would lead to a vertical upward motion of the wave. These results are in accordance with the observations and the obtained initiation centers on opposite sites of the active region in weak magnetic field. Simultaneous imaging from two different vantage points enabled us to apply the forward fitting model by \cite{thernisien06} and \cite{thernisien09}. From this we derived the flux rope position on the solar surface and found that it is close to the coronal dimming regions and wave initiation centers, respectively. Theses findings suggest that the CME expanding flanks are initiating the wave. A recent study by \cite{patsourakos10} showed that CMEs undergo a strong but short lived lateral overexpansion in their early evolution.

The mean speed of the wave of $\sim$240~km~s$^{-1}$, which is in the range of the characteristic speed of the ambient quiet solar corona, is derived to be similar for the western and eastern propagation direction as well as from observations of two different vantage points. The wave shows no clear deceleration which is due to its slow initial speed and, as a consequence of the extreme solar minimum, due to the weak magnetic field environment in which the wave is propagating \citep[see e.g.][]{kienreich11}. The derived speed of the wave lies within the velocity range of of 210--350~km~s$^{-1}$ as expected for fast magnetosonic waves during quiet Sun conditions \citep[see][]{mann99}. The radial speed of the upward moving CME is much larger and of $\sim$750~km~s$^{-1}$. In this respect it would be interesting to do a comparison to the model by \cite{attrill07} who interprets EUV waves as magnetic footprints of CMEs. This model would require a motion of the EUV wave similar to the global average motion defined by the lateral expansion of the CME. However, the CME is observed off-limb well after the end of the EUV wave and therefore it is not possible to compare simultaneous measurements of the wave and CME widths.

For the western propagation direction, the measurements from the 195~{\AA} passband show systematic larger distances as compared to 171~{\AA}. \cite{patsourakos09b} also found differences in the wave appearance between 171~{\AA} and 195~{\AA}, especially in the beginning of the wave evolution. For the eastern propagation direction the wave is observed to be more intense. The transformation of EUVI images from STEREO-B view to STEREO-A view makes foreshortening effects clearly visible (see Figure~\ref{3d}). Reconstructing the 3D structure of the wave by applying simple geometry, we find that the wave is inclined towards East. These results indicate that the 3D structure of the wave is not symmetric. From two independent methods (forward fitting model \cite{thernisien06} and triangulation \cite{temmer09b}) we derived that the propagation direction of the associated CME deviates eastward from the AR position. The appearance of the wave, especially its asymmetric intensity and 3D geometry, seems to be closely related to the non-radial evolution of the associated CME. The results from the present study support the wave nature of the phenomenon as well as the close association of the wave structure to the evolutionary characteristics of the expanding CME flanks.

Since in the EUVI 171~{\AA} passband wave features might be confused with expanding loops \citep{patsourakos09}, we would like to emphasize that the double-initiation center is found also from wave fronts identified in 195~{\AA} images. Moreover, the smooth kinematics observed in the eastern as well as western propagation direction gives no indication that the first fronts identified would not be associated to the propagating wave. In this respect we would also like to note that, if the first measured fronts are expanding loops, we face a very smooth transition between loop expansion and the build up of the wave. We expect that detailed studies with high spatial and temporal resolution SDO data will shed more light on the initiation phase of CMEs and their association to coronal waves. \cite{liu10} presented the first results from SDO imagery for the 8 April 2010 EUV wave and proposed a hybrid hypothesis combining both wave and non-wave aspects to explain the observations.

 \begin{acks}
We thank the referee for helpful comments. M.T. and A.V. gratefully acknowledge the Austrian Science Fund (FWF): P20867-N16.
 \end{acks}

%
%
%
\bibliographystyle{spr-mp-sola-cnd}

\end{article}
\end{document}